\documentstyle[prb,multicol,aps,psfig]{revtex}
\newcommand{\bfr}{{\mathbf r}}
\newcommand{\bfk}{{\mathbf k}}
\newcommand{\bfq}{{\mathbf q}}

\newcommand{\bfn}{{\mathbf n}}
\newcommand{\calH}{{\cal H}}
\newcommand{\bft}{{\mathbf \theta}}
\newcommand{\xhat}{  \widehat{{\bf x}}}
\newcommand{\be}{\begin{equation}}
\newcommand{\ee}{\end{equation}}

\begin{document}
\title{Twist-averaged Boundary Conditions in Continuum Quantum Monte Carlo}
\author{C. Lin, F.-H. Zong and D. M. Ceperley\\}
\address{Dept. of Physics and NCSA, University of Illinois at
Urbana-Champaign, Urbana, IL 61801}
\maketitle

\begin{abstract}
We develop and test Quantum Monte Carlo algorithms which use
a``twist'' or a phase in the wave function for fermions in
periodic boundary conditions. For metallic systems, averaging over
the twist results in faster convergence to the thermodynamic limit
than periodic boundary conditions for properties involving the
kinetic energy with the same computational complexity. We
determine exponents for the rate of convergence to the
thermodynamic limit for the components of the energy of coulomb
systems. We show results with twist averaged variational Monte
Carlo on free particles, the Stoner model and the electron gas
using Hartree-Fock, Slater-Jastrow, three-body and backflow
wavefunction. We also discuss the use of twist averaging in the
grand canonical ensemble, and numerical methods to accomplish the
twist averaging.
\end{abstract}

\vspace*{3mm} \noindent PACS Numbers: 02.70.-c, 82.20.Wt,
71.15.-m


\begin{multicols}{2}
\narrowtext

Almost all quantum Monte Carlo (QMC)calculations in periodic
boundary conditions have assumed that phase of the wavefunction
returns to the same value if a particle goes around the periodic
boundaries and returns to its original position. However, with
these boundary conditions, delocalized fermion systems converge
slowly to the thermodynamic limit because of shell effects in the
filling of single particle states. In this paper we explore an
alternative boundary condition: one can allow particles to pick up
a phase when they wrap around the periodic boundaries,
\begin{equation}
\Psi(\bfr_1 +L \xhat, \bfr_2, \cdots )= e^{i\theta_x} \Psi(\bfr_1,
\bfr_2, \cdots ).\label{TBC}
\end{equation}
The boundary condition $\theta = 0$ is called periodic boundary
conditions (PBC), $\theta = \pi$  anti-periodic boundary
conditions (ABC) and the general condition with $\theta \neq 0$,
twisted boundary conditions (TBC)\cite{twist}.

In periodic boundary conditions, the Hamiltonian is invariant with
respect to translating any particle around the periodic
boundaries. According to Bloch's theorem, this implies that any
solution can be characterized by a given twist angle. The twist
angle also has a physical origin: consider a toroidal geometry.
One can either rotate the torus\cite{leggett} and go into rotating
coordinates, or add a magnetic flux\cite{by} to the center of the
torus. The physical properties will be unchanged. In both cases
one can transform away the perturbation by applying TBC with the
twist angle given by $\theta=m R^2\omega/ h$ for rotation and
$\theta  = e\phi/(c\hbar)$ for magnetic flux.   A torus is
topologically equivalent to periodic boundary conditions, so that
a non-zero twist will be allowed in periodic boundaries. The twist
is a degree of freedom, or boundary condition, that can be varied
to enable a finite system to approach the thermodynamic limit more
quickly or to make detailed studies of the properties of the
quantum state.

If the periodic boundaries are used in all three directions, each
dimension can have an independent twist\cite{multitwist}. Hence,
in 3D, the twist is a three component vector, $\theta_i$ with
$i=\{1,2,3\}$. The free energy and hence all equilibrium
properties are (triply) periodic\cite{by} in the twist:
$F(\theta_i + 2 \pi )= F(\theta_i)$ so that each component of the
twist can be restricted to be in the range:
 \be  -\pi < \theta_i  \leq
\pi. \label{unitcell}
 \ee
For systems with a real potential ({\it e.g.} no magnetic field),
one can further restrict the twist to be in the range $[0,\pi]$.

For a degenerate Fermi liquid, finite-size shell effects are much
reduced if the twist angle is averaged over. We call this twist
averaged boundary conditions (TABC)\cite{TABCnote}. This is
particularly important in computing properties that are sensitive
to the single particle energies such as the kinetic energy and the
magnetic susceptibility. By reducing shell effects, much more
accurate estimations of the thermodynamic limit of these
properties can be made. What makes this even more important is
that the most accurate quantum methods have computational demands
which increase rapidly with the number of fermions. Examples of
such methods are exact diagonalization\cite{poilblanc}
(exponential increase in CPU time with N), variational Monte
Carlo\cite{cck}(VMC) with wavefunctions having backflow and
three-body terms\cite{kwon2d,kwon3d} (increases as $N^4$), and
transient-estimate and released-node Diffusion Monte Carlo
methods\cite{cep84} (exponential increase with N). Methods which
can extrapolate more rapidly to the thermodynamic limit are
crucial in obtaining high accuracy. Twist averaging is especially
advantageous for stochastic methods ({\it i.e.}  QMC) because the
averaging does not necessarily slow down the evaluation of
averages, except for the necessity of doing complex rather than
real arithmetic.

The use of twisted boundary conditions is commonplace for the
solution of the band structure problem for a periodic solid. Band
structure methods begin by assuming the wavefunction factors into
single particle orbitals characterized by a lattice momentum. Then
in order to calculate properties of an infinite periodic solid,
properties must be averaged by integrating over the first
Brillouin zone. Baldereschi\cite{balder} pointed out that in an
insulator, in integrating over the Brilliouin zone, one can with
high accuracy replace the integral with a ``special k-point.''
This was generalized to a grid of k-points\cite{monkhorst}.
Twisted boundary conditions has been discussed in connection with
polarization of insulators\cite{ivo}; we do not consider that
here. The use of twisted boundary conditions is common in the
analysis of lattice models\cite{poilblanc,julien}.
Gammel\cite{gammel} showed using perturbation arguments for
certain lattice models why it will converge faster to the
thermodynamic limit and applied it to calculating optical
properties. Gros\cite{gros} studied size effects in the Hubbard
model with exact diagonalization and showed TABC gives exact
results in the grand canonical ensemble for non-interacting
systems.

Though twisted boundary conditions have a long history within
quantum physics, their use in quantum Monte Carlo (QMC) has been
limited. In continuum QMC, Rajagopal\cite{raja} used special
k-points to reduce finite size errors for calculations of
insulators within VMC. Their use in Diffusion Monte Carlo (DMC)
results were restricted to PBC and ABC in order to work with real
wavefunctions. Kralik {\it et al.}\cite{louie} used generalized
boundary conditions to compute the momentum distribution  with VMC
for silicon (a semiconductor) and Filippi and
Ceperley\cite{filippi} for lithium (a metal). This was done in
order to enlarge the number of momentum vectors to probe the
behavior near the fermi surface. The use of TABC within QMC has
not been further developed.

The applications considered here are for systems with a fermi
surface, {\it i.e.} metals. We begin by discussing the method for
non-interacting (NI) fermions. Fermi liquid theory asserts that
the spectrum of states of interacting liquids are intimately
related to those of the non-interacting fermion systems, hence a
detailed analysis for the NI system carries over to strongly
interacting fermi liquids. We then discuss interacting systems in
the Hartree-Fock approximation: the electron gas and the Stoner
model. In the Stoner model, we show how TABC can be used to
determine a polarization phase transition at non-zero temperature.
Results for TABC are given for the interacting electron gas using
a pair product and backflow wavefunction in 3d. The electron gas
system has been previously treated with an extrapolation method
based on Fermi liquid theory. We show that TABC gives the same
results in the thermodynamic limit and verify the applicability of
the NI analysis, in particular to examine how the energy depends
on the twist of a given system size. We then present VMC results
of the polarization energy of the electron gas using the new
method. In future publications we will study the low density
properties of the electron gas using this technique. The Appendix
discusses some details arising in the implementation of TABC in
QMC.

\section{Non-interacting Fermions}

\begin{figure}
\centerline{\psfig{figure=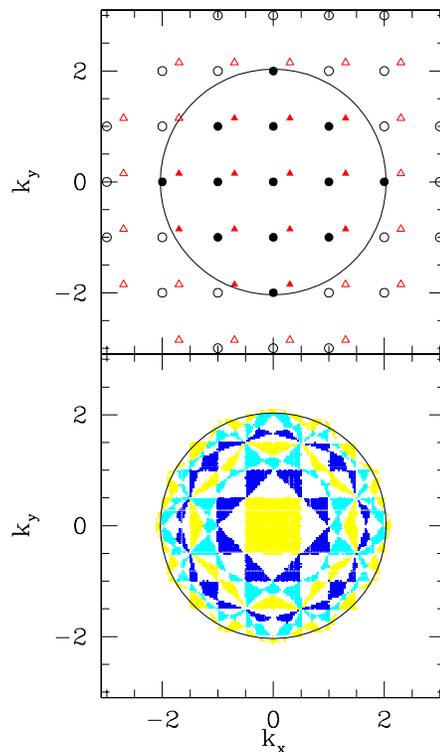,width=10cm,angle=0}}
\caption{Momentum distibution for 13 spinless fermions in a 2D
square with side $L=2\pi$. The top panel shows the occupied states
(closed symbols) and empty states (open symbols) with zero twist
(circles, PBC) and a twist equal to $2\pi (0.3,0.15)$ (triangles).
The circle shows the infinite system fermi surface. The bottom
panel shows the occupied states with TABC. The colored regions
show the occupied region for the lowest level (middle square), the
third level, up to the outermost 13$^{th}$ level.}
\label{occstates}
\end{figure}

In a non-interacting homogenous system with PBC, the single
particle states are plane waves: $\exp( i \bfk \bfr)\eta(\sigma)$
where $\eta$ is a spin function. For simplicity, we always assume
the simulation cell is a cube (or square in 2D) of side $L$. To
satisfy the twisted boundary conditions, the wave vectors obey:
\be \bfk_{\bfn} = (2 \pi \bfn +\bft)/L\label{momstates}\ee where
$\bfn$ is an integer vector. These states have energy
$E_{\bfn}=(\hbar^2/2m) \bfk_{\bfn}^2$. The ground state in the
canonical ensemble consists of the $N$ lowest energy states; the
many-body wavefunction is a determinant of those states.  In this
section we will ignore spin, since for a non-interacting system,
spin modifies the results only by doubling the degeneracy of each
level. Figure \ref{occstates} shows the occupation of states for
13 spin-less fermions in 2D for $\theta=0$ {\it i. e.} with
periodic boundary conditions, and also with a non-zero twist. The
occupied states lie within a circle centered at the origin with
radius $\approx k_F=2 (\pi \rho)^{1/2}$.

\begin{figure}
\centerline{\psfig{figure=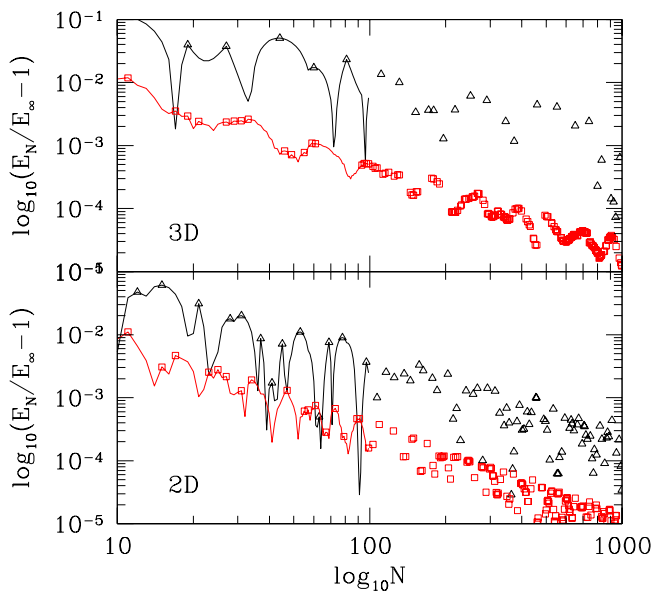,width=10cm,angle=0}}
\caption{Relative error of the energy  versus number of particles
with PBC ($\triangle$) and TABC ($\Box$) in 2D and 3D. The points
shown are only those where the relative error has a local maximum.
Curves are shown only for $N<100$. }\label{decay}
\end{figure}

Figure \ref{decay} shows the relative error in energy versus the
number of fermions with PBC. The energy converges slowly to the
exact result. One sees ``cusps'' in the curve at certain values of
$N$. These occur at closed-shell values of $N$, {\it e.g.} the
state depicted in Fig. \ref{occstates} for PBC is a closed shell
since states related by symmetry are either all filled or all
empty. For large $N$ the curve is ``quasi-random'', with an
envelope decaying algebraically as $ N^{-\nu}$.

We find numerically that the exponent of the decay of the relative
error of the energy is approximately  $ \nu = 1.33$ in 2D and $\nu
= 1 $ in 3D (see Table I). To characterize the approach to the
thermodynamic limit, we introduce two different measures. Defining
the relative scaled error $\delta_N$ as: \be E_N = E_{\infty}( 1+
N^{-\nu} \delta_N),\label{fdecay} \ee we define $a=\max
|\delta_N|$, $b=\langle\delta_N\rangle$ and
$c=\langle(\delta_N-b)^2\rangle^{1/2}$. Table I shows estimates of
these coefficients and exponents obtained numerically by examining
values of $10 \leq N \leq 10^4$.

Now consider twisting the boundary conditions, {\it i. e.} using a
nonzero phase. This displaces the set of k-vectors as shown in the
top panel of Fig. \ref{occstates}. Aside from a set of special
twists having zero measure, the energy levels will no longer be
degenerate. (When we sort the states to decide the filling, all
states will have a different energy.) This is because inversion
symmetry and rotational symmetry through 90$^{\circ}$ is broken.
This breaking of symmetries and absence of degeneracies is a
crucial difference between TBC and PBC.

\begin{figure}
\centerline{\psfig{figure=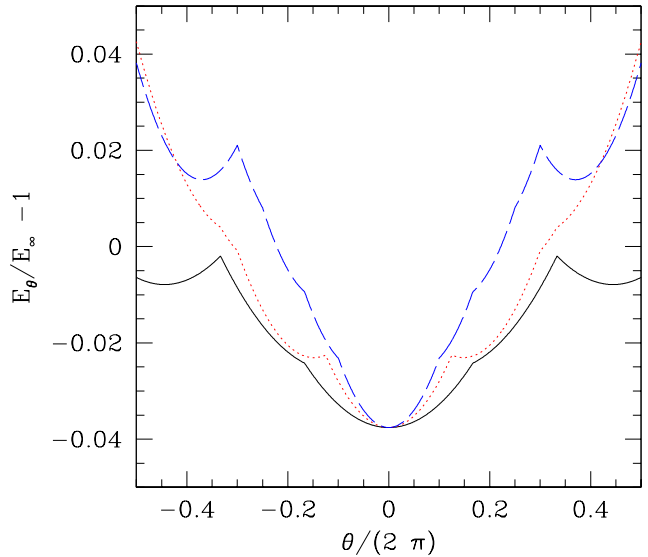,width=10cm,angle=0}}
\caption{Dependence of the energy  of NI unpolarized fermions on
the twist angle for $N$=54 in 3D. The solid line shows the energy
along the (100) direction, dotted line along the (110) direction
and dashed line along the (111) direction. The curves are
piecewise quadratic, with a cusp when the occupation of the states
changes. We refer to the r.m.s. spread of energy values as the
bandwidth. }\label{bande}
\end{figure}

At critical values of the twist, when a filled and empty state
have the same energy, the occupation of the states changes. The
condition for the degeneracy is that $\bfk$ lie on a plane
bisecting and perpendicular to the line joining the origin with an
integer vector; precisely the Laue condition for the Bragg
planes\cite{ashcroft} of the reciprocal lattice of the supercell.
In fig. \ref{bande} is shown the dependence of total energy on the
twist angle for a fixed number of particles. One sees cusps as the
filled states cross the Bragg planes. The dependence is similar to
the band energy of a simple metal.  Later, we will discuss this
band structure for an interacting system. The bandwidth $E_{BW}$
(the spread of energy values fig. \ref{bande}) is defined as: \be
E_{BW}^2= (2\pi)^{-d}\int d\bft
(E(\bft)-E_{\infty})^2\label{bandwidth}\ee depends on the number
of particles and scales as $E_{BW} \propto N^{-\nu}$ where the
exponent is the same as describes the convergence of the KE in
PBC.

There are several alternative procedures by which the twist angle
can be varied: i) one can average the twist over all possible
values, ii) the twist can become a dynamical variable and iii)
special values of the twist could be used. Of these approaches,
none are right or wrong in general; which method approaches the
thermodynamic limit faster depends on the order of the phase in
question, whether fermi liquid, ferromagnetic or
anti-ferromagnetic. However, to compute a variety of properties
for a metallic systems, we find the TABC best reduces size
effects.

\begin{table}
\begin{tabular}{|c|c|c|c|c|c|c|}\hline
 n & O & d & $\nu$ & a &    b & c \\\hline
  1  & T & 2  &  1.33  &  4.5 & 0.37  & 1.77 \\
  8  & T & 2  &  1.5  & 0.47 & 0.27  & 0.093 \\\hline
  1  & N & 2   & 0.67  &  2.18    &   0   &   0.65    \\
  16  & N & 2   & 0.75  &  0.71    &   0      & 0.52      \\\hline
  1  & V & 2   & 1  &   0.50   &   -0.35      & 0.069      \\
  8  & V & 2   & 1  &   0.38   &   -0.367      & 0.0058      \\
  \hline
  1  & T & 3  &   1   & 2.4  & 0.25  & 1.0 \\
  8  & T & 3  &  1.33  & 0.50 & 0.292  & 0.065\\
 16  & T & 3  &  1.33  & 0.35 & 0.21  & 0.06\\
 32  & T & 3  &  1.33  & 0.35 & 0.19  & 0.06 \\\hline
 1   & N & 3   & 0.55  &   2.97   &  0    &   1.00    \\
 16  & N & 3   & 0.67  &  0.83   &   0     &  0.63    \\\hline
 1   & V & 3   & 0.67  &   0.742   &   -0.549      & 0.072      \\
 16  & V & 3   & 0.67  &   0.587   &   -0.582      & 0.0043      \\
 \hline
\end{tabular}
\caption{Coefficients of the asymptotic decay of the error in the
relative NI energy. $\nu$ is the exponent of the decay. The
exponents have been determined from the numerical data and are
accurate to about 0.02. The amplitude was determined numerically
by examining the values for $10 \leq N \leq 10000$. The
coefficients are defined as $a\equiv\max(|\delta_N|)$, $b= \langle
\delta_N \rangle$, $c= \langle [ \delta_N -b]^2\rangle^{1/2}$. $n$
is the number of phase angles used for the summation in each
dimension: $\theta_i = 2 \pi i /n$ for $i = 1, \ldots , n$. $n=1$
corresponds to PBC; d is the dimensionality. $O$ is the property:
$T$, the kinetic energy; $V$, the Hartree-Fock potential energy of
the electron gas; $N$, the number of particles in the TA-GCE
method. }
\end{table}

\section{Twist averaging}

The twist average of a property $\widehat{A}$ is defined by:
 \be
\langle \widehat{A} \rangle = (2 \pi)^{-d} \int_{-\pi}^{\pi} d\bft
\langle \psi(R,\bft) | \widehat{A}| \psi(R,\bft)\rangle
 \ee
where it is assumed that the wavefunction $\psi(R,\bft)$ is
normalized for each $\bft$. The momentum distribution, $n(\bfk)$
is a key property to calculate for delocalized quantum systems.  A
discontinuity in $n(\bfk)$ at the Fermi surface is responsible for
the validity of Fermi Liquid Theory for metals. The kinetic energy
is the second moment of the momentum distribution: \be T_N =
(\hbar^2/2m) \int d\bfk \bfk^2 n(\bfk).\ee

Let us analyze the momentum distribution for NI fermions in the
canonical ensemble. For any given twist $\bft$, the $N$ lowest
energy states from those given by Eq. \ref{momstates} are
occupied. But any value of $\bfk$ can only be reached by a unique
combination of $(\bfn , \bft)$ if $\bft$ is restricted by Eq.
\ref{unitcell}. This proves that the averaged momentum
distribution is a constant for states which can be reached by some
combination of $(\bfn , \bft)$ and zero otherwise. Hence within
TABC, the set of filled states comprises a ``solid volume''
bounded by a Fermi surface. In contrast, for a single twist value,
the momentum distribution is a point set. The total volume in
k-space inside the Fermi surface is precisely $(2\pi)^d\rho$, just
as it is in the thermodynamic limit, so the constant is determined
by the normalization condition: \be \int d\bfk n(\bfk) = 1. \ee As
mentioned above, the Fermi surface is a subset of the Bragg
planes. For $N$ particles the occupied states comprise the union
of the  first $N$ Brillouin zones\cite{ashcroft}. The $(N+1)^{th}$
electron will go in the $(N+1)^{th}$ zone, an area formed by
planes surrounding the $N^{th}$ zone. Figure \ref{occstates} shows
the momentum distribution of 13 spinless fermions in 2D using
TABC.

In 1D, TABC gives the exact momentum distribution because the
normalization condition determines everything. In higher
dimensions the fermi surface is not perfectly circular (spherical)
as shown in fig. \ref{occstates}. However, one can see that
$n(\bfk)$ is much closer to a disk than the momentum distribution
obtained with PBC. A perfect  fermi surface (no finite size
corrections) in any dimension, can be obtained by allowing
variations in the particle number and working in the grand
canonical ensemble as we discuss below.

Figure \ref{decay} shows the convergence of the error of the
kinetic energy within TABC versus the number of particles.
Numerical estimates of the relative error are given in Table I.
One sees a dramatic improvement in the convergence with respect to
PBC. The exponents governing the decay rate are larger and the
errors are up to two orders of magnitude smaller for system sizes
in the range $N\approx 100$. Note that the TABC kinetic energy
must approach the exact energy from above. This is because the
shape of a given volume with the smallest moment of inertia is a
sphere, so that the distorted shape shown in fig. \ref{occstates}
has a higher energy. Also shown in Table I is dependence of the
error on the number of twist values in the average. One needs from
16 to 32 values of $\theta$ along each axis to achieve the full
reduction in size effects (better than a percent accuracy in the
relative error of the size effects). In the Appendix are discussed
the relative merits of performing the average on a grid versus
sampling the twist values from a uniform distribution.



Let us now examine how the potential energy converges with PBC and
TABC. This will give us some idea of how two particle correlations
are affected by the boundary conditions since the potential energy
is a particular integral over the pair correlation function. The
calculation performed below is particularly simple for a power law
potential, $v(r)=r^{-n}$. In particular, we examine the potential
energy of an electron gas $(n=1)$ computed using the NI
wavefunction (Hartree-Fock approximation). The NI trial function
is valid for high density when the kinetic energy dominates the
potential energy. The potential energy (using the Ewald image
potential) is conveniently evaluated in Fourier space as a sum
over the structure factor:
 \be V =\frac{N\rho}{2} \sum_{\bfk} v_{\bfk}
(S_{\bfk}-1) + Nv_M
 \ee
where $v_M$ is the Madelung energy of a charge interacting with
itself and $ v_{\bfk}$ is the Fourier transform of the
interparticle potential. For a $1/r$ potential, $v_k =
2\pi(d-1)/k^{d-1}$. The values of $\bfk$ in the sum are $\bf k =
2\pi \bfn / L$ where $\bfn$ is an integer vector. For the NI
wavefunction, the structure factor at wave-vector $\bfq$ is
proportional to the probability that after we have displaced a
filled state by $\bfq$ we are still in a filled state. \be
S_{\bfq} = 1-\frac{1}{N} \langle \sum_{\bfk < \bfk'}
\delta(\bfk-\bfk' -\bfq)\rangle
 \ee
where the sum is over occupied states and the average is over
twisted boundary conditions.

\begin{figure}
\centerline{\psfig{figure=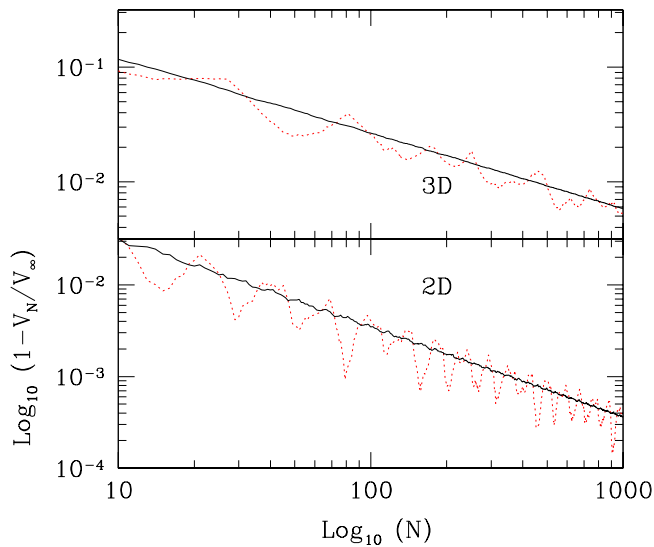,width=10cm,angle=0}}
\caption{Relative error in the evaluation of the potential energy
for an electron gas using the Hartree-Fock wavefunction for $N$
spinless electrons.  The solid line shows TABC, the dashed line
PBC. }\label{vdecay}
\end{figure}

Shown in fig. \ref{vdecay} is the convergence of the potential
energy versus the number of particles using PBC and TABC. For all
values of $N$ and twists the potential energy of the finite
systems approaches that of the infinite systems from below. Twist
averaging serves to make the decay more regular but does not
reduce its overall magnitude which is determined by a charge
interacting with the correlation hole in the nearby supercell.
Similar effects are expected for other two particle quantities.
The smoother convergence obtained with TABC should allow for more
accurate extrapolation to $N\rightarrow\infty$.

\section{Grand Canonical Ensemble}

The use of TABC in the grand canonical ensemble (GCE) gives the
exact single particle occupations for NI particles, as shown by
Gros\cite{gros} within the Hubbard model. Suppose $(\alpha,N)$ is
a label of the quantum states, both for the number of particles
and for other quantum numbers such as the momentum and let
$E_{\alpha,N}(\bft)$ be the energy of this state. Then the
probability of a given state in the GCE is proportional to
$\exp(-\beta(E_{\alpha,N}(\bft)-N\mu))$ where $\mu$ is the
chemical potential. In the ground state, $\beta
\rightarrow\infty$, the occupied many-body state will be the one
minimizing $E_{\alpha,N}(\bft)- N\mu$. Thus $N$ can depend on
$\bft$.


We now show that for a NI system, TABC within the GCE gives exact
single particle properties; {\it i.e.} there are no finite size
effects. Suppose the single particle energy levels are $e_k$. Then
the probability of occupying the $N$ states $e_1, \ldots , e_n$ is
$\exp(-\sum_{k=1}^N \beta[e_k(\bft)-\mu])$. In the occupation
number $\widehat{n}_k$ basis, this probability distribution
factorizes as: \be \prod_{\bfk} \left[\widehat{n}_k
e^{-\beta[e_k(\bft)-\mu]} + (1- \widehat{n}_k)\right]\ee so the
probability of state $k$ being occupied is precisely the fermi
distribution law $n_k=(\exp(\beta(e_k-\mu))+1)^{-1}$. As the twist
angle is varied over its range, each momentum state of the
infinite system occurs precisely once. Hence the averaged
occupation number is precisely what it would be in the
thermodynamic limit. The distortion of the fermi surfaces observed
in the lower panel of fig. \ref{occstates} is a consequence of
using the canonical ensemble.

The momentum distribution and hence the kinetic energy will be
exactly equal to their infinite system values. Other properties
may have finite size corrections, only the single particle
properties are guaranteed to be exact. We call this procedure the
twist average grand canonical ensemble (TA-GCE).

With this procedure one does not have a fixed number of particles
since for a given twist and fermi wave vector, the number of
occupied states will vary. The fluctuations in the number of
particles is closely related to a famous problem in analytic
number theory, ``Gauss' circle problem'', to determine the number
of lattice points inside a circle of area A as its radius tends to
infinity\cite{mathency}. As Gauss posed the problem, the center of
the circle was fixed on a lattice site while in TA-GCE the circle
is placed randomly (the shifted circle problem).   It has been
shown\cite{kendall} that in 2D one obtains the following
fluctuation in the particle number for TA-GCE: \be
\lim_{N\rightarrow \infty }\langle (N- (Lk_F)^2/4\pi)^2
\rangle^{1/2} \propto N^{1/4}. \ee Our numerical estimates for the
convergence are shown in Table I. Note that the Gauss circle
problem differs from the convergence of the energy in the
canonical ensemble since the energy is the average second moment
of the lowest $N$ points, not the number of points in a circle.

The preceding discussion refers to NI systems, however, fermi
liquid theory asserts that the low lying excited states of an
interacting system are in one-to-one correspondence with the NI
states. Hence, as argued by Gross\cite{gros} and
Gammel\cite{gammel}, TA-GCE is likely to reduce finite size
effects substantially for interacting systems.

For an interacting many-body system, one difficulty in using the
GCE is the need to optimize the wavefunction at each twist value;
in particular to pick out which orbitals should be occupied. For
an isotropic system having a spherical fermi surface, the order of
filling the states is simple.  For a metal with a non-spherical
Fermi surface, the usual procedure is to determine the filling of
single particle states according to a mean-field theory such as
the Kohn-Sham method in density functional theory. If one uses the
same procedure within QMC, the Fermi surface will be substantially
unchanged. Hence, it would be better to try other fillings,
choosing the one which minimizes $E_N-N\mu$.

There are other problems in using the GCE for charged systems in
periodic boundary conditions. Usually the positive compensating
charge, either a uniform background or a fixed array of charged
nuclei, is at a fixed density. But if the number of electrons
fluctuates, the periodic cell can have a net charge, which causes
problems in calculating the
long-range potential. 
Although very promising, we do not consider the TA-GCE method
further in this paper. The following examples are for the
canonical ensemble.

\section{The Stoner model}

To test the utility of TABC for determining a phase transition, we
simulated the Stoner model, an analytically soluable
model\cite{stoner,suris} for strongly correlated fermions,
particularly related to itinerant magnetism\cite{herring}. The
Stoner model differs from NI fermions by the addition of a contact
repulsive potential: $\sum_{i<j} g \delta(r_{ij})$. The energy is
evaluated within the mean field (Hartree-Fock) approximation using
the NI wavefunction. For a spin $(1/2)$ system, the potential
energy is $E = E_{NI} + g n_\uparrow n_\downarrow$.  In 3D the
energy at zero temperature in the thermodynamic limit is: \be E/N
= \lambda \frac{(3 \pi^2 \rho )^{5/3} }{10 \pi^2} [ (1 + \zeta
)^{5/3} +(1-\zeta)^{5/3} + b (1 - \zeta^2) ] \ee where $b =
\frac{5 g \rho}{6 \lambda} (3 \pi^2 \rho)^{-2/3}$ and
$\lambda=\hbar^2/2m.$ For $b< 1.111$ the system has an unpolarized
ground state and for $b>1.3228$ the ground state is ferromagnetic.
For intermediate couplings, the ground state has a partial spin
polarization at zero temperature, similar to the behavior
suggested\cite{acp82,ortiz} for the electron gas at low density.

\begin{figure}
\centerline{\psfig{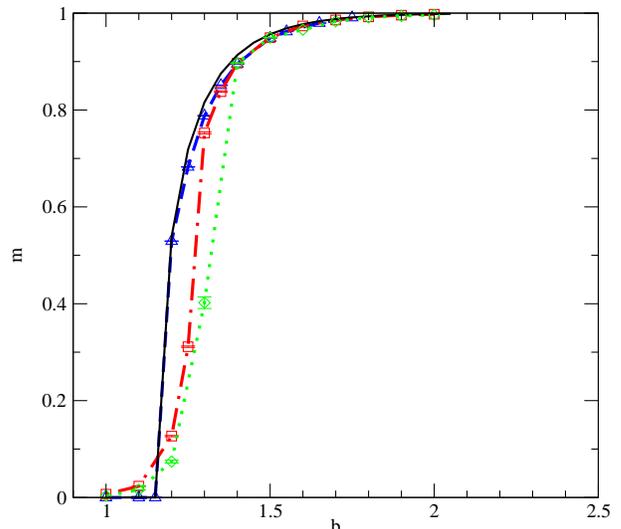}}
\caption{Magnetization as a function of $b$ for the 3D Stoner
model. Circles are the exact results, diamonds for 54 electrons
with PBC, squares for 54 electrons with TABC and triangles for
N=200 with TABC.  The temperature was $T=0.224E_F$ and a $8^3$
grid of twist values was used. }\label{stoner}
\end{figure}

We performed a MC simulation of the 3D NI system, within the
occupation representation. The state variables of the system
consist of occupation numbers (both spin and wavevector) and the
twist angles. We make Monte Carlo moves consisting of flipping
spins and moving the spatial occupation while keeping the particle
number fixed. If the new state is not already occupied, the move
was accepted with probability equal to $\exp (-\beta (e_{new}
(\bft) - e_{old} (\bft) ))$. Fig. \ref{stoner} shows the
convergence of magnetization distribution versus the number of
fermions. We found that one could determine the phase transition
within a few percent accuracy in the coupling constant $b$ using
only 54 fermions. Such accuracy within PBC would require thousands
of fermions because of the strong shell effects. \vskip 1cm

\section{Electron gas}

We now present results for the 3-D electron gas as a test of TABC
on a correlated many-body continuum system.  The electron gas is a
very important model in condensed matter physics being the basis
for the density functional theory method of electronic structure
computations.\cite{cep80} The phase diagram of the electron gas at
low density is still not resolved\cite{ortiz}. The wavefunctions
we use are the most accurate known for a continuum many-body
system with optimized 2-body, 3-body and back flow terms.
Variational MC (VMC) and Diffusion (DMC) are QMC methods
appropriate to zero temperature, and Path Integral (PIMC) to
$T>0$. In this paper, we will only discuss VMC and DMC. PIMC will
be discussed in future publications.

In VMC, one assumes an analytic form for a trial function $\Psi_T
(R)$ where $R$ symbolizes the $3N$ coordinates.  Then one samples
$|\Psi_T (R)|^2$ using a random walk\cite{cck}. An upper bound
estimate to the exact ground state energy is the average of the
local energy $E_L (R) = \Psi_T (R) ^{-1} \calH \Psi_T (R)$ over
the random walk. An accurate trial wavefunction is obtained from
the NI wavefunction by multiplying by pair correlation terms. The
pair product  or Slater-Jastrow wave function is: \be \Psi_2 (R) =
Det (e^{i\bfk_i \bfr_j}) e^{-\sum_{i<j} u(r_{ij})}. \ee

\vskip 1cm
\begin{figure}
\centerline{\psfig{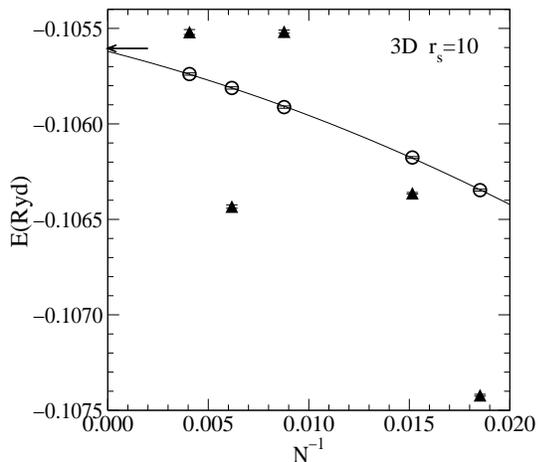}}
\caption{Energy versus the number of electrons for the 3D electron
gas using SJ wavefunction. The circles and connecting line are
TABC with 10$^3$ twist values. The filled $\triangle$'s are using
PBC. The arrow shows the extrapolation to an infinite system made
using the PBC calculations using Eq. \protect\ref{extrapolation}.
\label{3DEng}}
\end{figure}

One can determine the correlation factor $u(r)$ either with an
analytic argument or by a minimizing of the energy and/or the
variance of an assumed form. We use a parameter-free analytic form
so that the systematic size and twist effects are not masked by
noise in the trial function itself. For the electron gas very
accurate analytic form for $u(r)$ based on the energy within the
RPA approximation\cite{cep78} has as low an energy as those with
optimized parameters. We used optimized Ewald sums \cite{natoli}
both for the potential and for the correlation factor so as to
have the correct long wavelength behavior.


Fig. \ref{3DEng} shows the convergence of the VMC energy versus
the number of particles within TABC and PBC.  One can see the
convergence to the thermodynamic limit that was found within the
NI system is also evident within VMC using the SJ wavefunction. In
addition to the Slater-Jastrow trial functions, we also have used
optimized backflow-three body functions(BF3B)\cite{kwon3d}  which
give a more accurate description of the low-density electron gas.
(They pick up about 3/4 of the remaining correlation
energy\cite{kwon3d} and break certain symmetries of the SJ
wavefunctions.)

In the Diffusion Monte Carlo (DMC) method, one starts with a trial
function and uses $\exp(-t H)$ to project out the ground state
using a branching random walk. Fermi statistics pose a significant
problem to the projection method, since exact methods such as
transient estimate or release-node QMC suffer a exponential loss
of efficiency for large numbers of particles. For this reason the
approximate fixed-node method is normally used.  Using the
fixed-node method, one obtains the best upper bound to the energy
consistent with an assumed sign of the wave function. Both the FN
method and the exact transient estimate can be generalized to
treat complex-valued trial functions.\cite{fp} These methods are
called the fixed-phase and released phase QMC. We have tried both
of these approaches using TABC.\cite{lin_thesis}

\section{Fermi Liquid Theory and TABC}

Fermi liquid theory (FLT) for metallic systems allows both a
method to extrapolate to the thermodynamic limit and a way of
understanding the twist dependence of the QMC results.  According
to Landau, the low-lying excitations of an interacting system are
in close relation to those of the NI system.

\be E = E_0 + \int d\bfk \delta n_{\bfk} e(\bfk) + \int d\bfk
d\bfk ' \delta n_{\bfk} \delta n_{\bfk'} f(\bfk,\bfk') + \dots
\label{flteq}\ee where $\delta n_{\bfk}$ is the deviation of the
quasi-particle occupation from the ground state, and $e( \bfk )$
and $f( \bfk , \bfk ')$ are 1 and 2 quasiparticle energy
functionals. For simplicity we have not indicated dependence on
spin. The energy functionals are usually further expanded about
the fermi surface in spherical harmonics and applied to calculate
properties in the thermodynamic limit. Here we discuss how to
apply FLT when the ``excitation'' is caused by the boundary
conditions. For example, we consider the momentum distribution
shown in fig. \ref{occstates}. The change in the energy caused by
the non-circular shape should also be given by Eq. \ref{flteq}.

\begin{figure}
\centerline{\psfig{figure=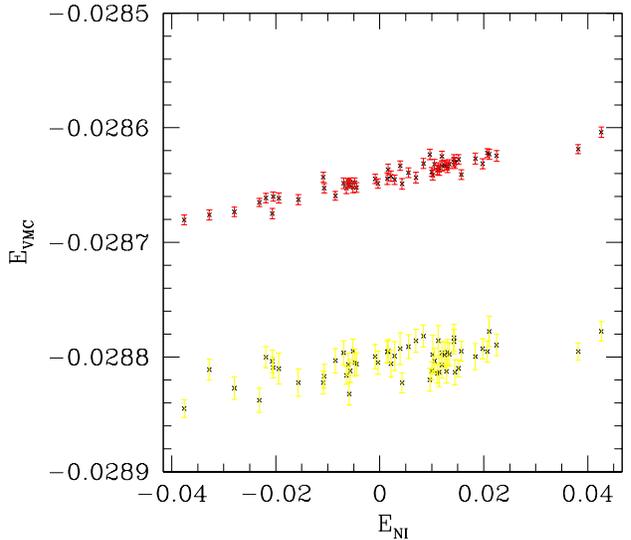,width=10cm,angle=0}}
\caption{Plot of the VMC energy versus the deviation of the NI
energy from $E_{\infty}$. Each phase angle is plotted separately.
Simulations are for $N=54$ unpolarized 3D electron gas at
$r_s=50$. Since the excitations are linearly related, Fermi liquid
theory describes the phase angle dependence.  Upper points are the
Slater-Jastrow wavefunction, lower ones are done with the
backflow-3-body trial function. Effective masses (the slope) are
respectively 1 and 0.61 for the two trial functions. }\label{flt}
\end{figure}

We can analyze this dependence by  comparing the energies of the
non-interacting system within a given twist with the interacting
system. This is done in Fig. \ref{flt}. One sees a linear relation
between the two energies, confirming that for these wavefunctions,
Fermi liquid theory is an appropriate description. The slope is
the effective mass of the quasi-particles. Then since the NI
infinite energy is known this gives a way of determining the
interacting system energy.

Previous calculations on the electron gas have used another
application of FLT, the extrapolation method\cite{cep78,hydrogen}.
In this method, one calculates accurate ground state energies for
a sequence of particle numbers, and determines the effective mass,
potential correction and infinite system energy by fitting these
energies to the relation: \be E_N=E_{\infty}+ (m/m^*) \Delta T_N +
\epsilon N^{-\nu}\label{extrapolation}\ee where $ \Delta T_N$ is
the deviation of the NI kinetic energy from the infinite system
and $\nu$ is the exponent for the potential energy given in Table
I. Figure \ref{3DEng} shows that the estimate of the infinite
system energy obtained using TABC and the extrapolation method
agree within errors. This is reassuring, but expected, since both
are based on FLT.

However, applied in practice, there are significant advantages to
the TABC method. The fitting method requires well converged runs
for at least three different values of $N$.  This can be
difficult, for example when the unit cell is large. For example,
suppose one is doing a simulation of bcc hydrogen where one is
limited to values of $N$ equal to $2K^3 = \{ 2,16,54,128,250
\ldots \}$. The last 3 values are reasonable to simulate, however,
if the unit cell contained 10 electrons, for example, the
extrapolation, would be very difficult without further
approximations. There is also the problem that convergence and
trial functions for large $N$ may differ from that for small $N$.
For example, one typically determines $m/m^*$ and $\epsilon$
within VMC and then applies the extrapolation using the DMC
energies. There are other problems with extrapolation. For
example, suppose we have a partially spin polarized system with
say $n_1$ spin up particles and $n_2$ down particles. Using PBC
there are no necessarily larger, closed shell systems with
precisely the same ratio $\zeta=(n_1-n_2)/(n_1+n_2)$.

\begin{figure}
\centerline{\psfig{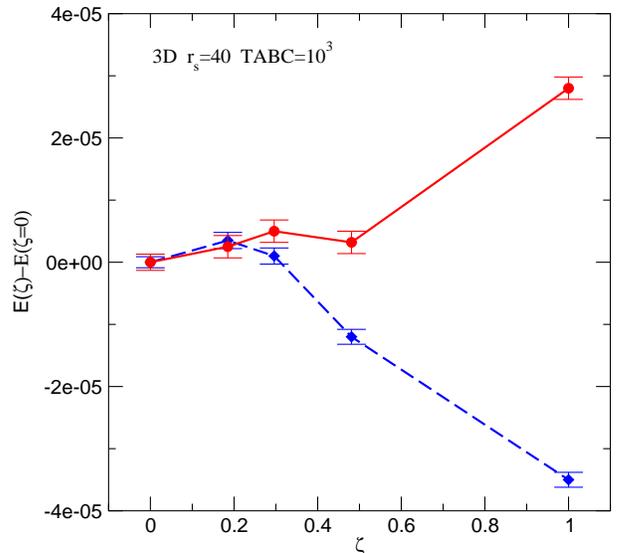}}
\caption{ Energy versus spin polarization for the 3D electron gas
at $r_s$=40 and $N=54$. The dashed line is with the Slater-Jastow
function, the solid line with backflow-3-body trial function.
Calculations are done using VMC with $10^3$ twist
values.}\label{polarization}
\end{figure}

Within TABC, there is the possibility of obtaining results in the
thermodynamic limit, using only quantities computed for a single
value of $N$. Figures \ref{flt}-\ref{polarization} demonstrate
that the kinetic contribution is well controlled.  Other
methods\cite{kent} are needed to make the potential correction
also for a fixed $N$. Use of TABC should make this more accurate
as well, as demonstrated in fig. \ref{vdecay}.

\section{Other procedures using twisted boundary conditions}

\noindent{\bf Dynamical twist method.} As an alternative to TABC,
one can let the twist angle be a dynamical variable (DTBC) and be
determined self consistently. By dynamical is meant that the
probability of a given twist is proportional to its free energy:
 \be  P(\bft) \propto \exp( -\beta F(\bft) ) = \sum_{\alpha} \exp (-\beta
E_{\alpha}(\bft) )
 \ee where $E_{\alpha}(\bft)$ is the energy  of the state $\alpha$ with
twist $\bft$.   This distribution of twist angles could be
attained within QMC by enforcing detailed balance on moves of the
twist angle during a random walk. The expectation of an operator
$A$ will be: \be \langle A \rangle_{DTBC} = \frac{1}{Z} \int d
\bft e^{-\beta F(\bft)} \langle \Psi_{\alpha} (\bft)| A(\bft)|
\Psi_{\alpha} (\bft) \rangle.
 \ee
At zero temperature, a special set of twist angles, those which
have the lowest energy, will be singled out, so that the ground
state energy will be $E_{DTBC} = \min_{\bft} (E(\bft))$. Clearly
$E_{DTBC} \leq E_{\infty}$, in contrast to the TABC which gives an
upper bound. In general, the dynamical energy, $E_{DT}$ converges
as slowly to the thermodynamic limit as does the PBC energy: the
exponents and fluctuations are the same. The dynamical twist
method is not an improvement over PBC for approaching the
thermodynamic limit for a metallic system at temperatures much
lower than the fermi temperatures.

Although the dynamic twist method is not satisfactory for a Fermi
liquid at low temperatures, for certain lattice models such as an
antiferromagnetic Heisenberg model, it can be a definite
improvement over PBC and TABC. This is because one can only
establish a defect-free N\'{e}el ordering if the unit cell is
commensurate with the boundary conditions. For the Heisenberg
model on a triangular lattice, the ground state has a given twist
per lattice spacing. Using DTBC allows one to establish this twist
value automatically, without imposing it in advance. This is
equivalent to the classical variable cell method where the
dimensions and aspect ratio of the supercell of a crystal become
dynamical variables, so that one can determine the most stable
crystal lattice structure\cite{RP} instead of examining each
crystal structure explicitly.

\vskip .6cm \noindent{\bf Special points} For each value of $N$
there exists a set of twist values for which $E(\bft)=E_{\infty}$.
One can determine these special twist values for the NI system and
then perform simulations at only one of those twist values for the
interacting system, thereby getting rid of single particle size
effects. This is similar to the special k-point method of
Baldereschi\cite{balder} for insulators where a single k-point is
determined by symmetry, thus allowing one to replace an integral
over the Brillouin zone with evaluation at a single k-point. This
method was used within QMC by Rajagopal et al. \cite{raja}.

The special k-point method is not appropriate for a metal because
of the discontinuity in properties at certain twist values. One
cannot replace the average by a single point because the Fermi
surface is not given in advance by symmetry, and can change
between the interacting and non-interacting wave functions. In
addition, it is not expected that the same twist values
appropriate for the NI energy, will be appropriate for other
quantities such as the potential energy, or spin susceptibility.
It is better to have a method which can give a spectrum of
properties correctly, rather than only the kinetic energy. As we
discuss in the Appendix, TABC does not impose an excessive
computational burden, and is to be preferred over using only
special twist values.

\section{Conclusions}

Note that there is a significant difference in the efficiency for
stochastic (QMC) methods versus explicit methods such as density
functional theory or exact diagonalization in regards to the TABC.
In explicit methods, computations for each twist require an equal
amount of computer time so that averaging over $N_G$ twist values
will take roughly $N_G$ times as long as a single twist value. One
can use inversion and rotational symmetry to reduce this, so that
for a grid of 16$^3$ points, TABC will require only $165$ twists.
On the other hand, in QMC all the twists reduce the statistical
error of the average. The twists are simply three more degrees of
freedom on top of the $3N$ coordinate variables to be averaged
over. However, one must also take into account startup costs
associated with each twist, such as re-optimizing the trial
wavefunction or equilibrating the random walk. Neglecting these
startup costs, there is no loss of statistical efficiency in
performing TABC so that the gain in reducing the systematic error
is free. This is examined in more detail in the Appendix.

There are many examples where twist averaging can effect
considerable improvement over the use of PBC. We have been able to
perform quite accurate calculations of the polarization energy of
the 2 and 3D electron gas\cite{lin_thesis} and of liquid $^3$He
using backflow wavefunctions with on the order of 100 fermions. As
pointed out by Ortiz et al.\cite{ortiz} calculations on such small
systems in PBC have considerable systematic errors. One property
that could be computed more accurately with TABC is the estimation
of fermi liquid parameters by calculating particle-hole
excitation\cite{kwon_flp}. TABC can reduce the shell effects which
caused much difficulty in that calculation. A related example is
in computation of the charge response of the electron gas
\cite{moroni} where a considerable effort was made to cancel out
effects of the PBC. We are presently studying the electron gas
confined to a slab\cite{slab} to determine the work function and
surface energy of a metallic surface. The filled states consist of
a set of disks, each of which will have a certain occupation
number. By doing twist averaging we have shown reduced size
effects with respect to PBC.

Experimental systems are at a non-zero temperature. For NI systems
one occupies the states with probability given by the Fermi-Dirac
distribution.  Because the Fermi function at non-zero temperature
is a continuous function, the convergence to the thermodynamic
limit will be much faster, even with PBC.  However in practice,
one is interested in electronic systems close to the ground state;
the relevant quantity is the thermal deBroglie wavelength of the
electron: $\hbar/(m_ek_bT)^{1/2} \approx 32 \AA$ at $T=300K$.
Because this length is usually larger than the simulation cell in
QMC, the localization of the density matrix does not help at
reducing fermion finite size at these temperatures. Using a
non-zero temperature just to achieve faster convergence to the
thermodynamic limit is not practically useful. However, TABC is
extendable to finite temperature PIMC simulations using the
fixed-phase method\cite{fp} and will reduce size effects at low
temperature.

The TABC method is likely to be valuable for all QMC calculations
in systems with a fermi surface. The calculations on the electron
gas demonstrate that even though it may be a little slower per
step of the random walk, it is better to do TABC than a larger
system with periodic boundary conditions because TABC converges
much faster to the thermodynamic limit. The overall efficiency of
any numerical method is ultimately judged by the computer time
needed to reduce systematic {\bf and} statistical error below a
given value. TABC is effective in reducing the systematic errors
and thus improve the overall efficiency.

This research was supported by NSF DMR-98-02373 and the Department
of Physics at the University of Illinois Urbana-Champaign. We
acknowledge useful discussions with R. M. Martin and G. Bauer.
Computational resources were provided by the NCSA. We thank H.
Edelsbrunner for references concerning Gauss' circle problem.

\section*{Appendix}

Here we discuss numerical details of implementing twist averaged
boundary conditions. Many of the changes caused by twisted
boundary conditions arise from the need to have complex
wavefunctions. Although the wavefunction and energies in special
cases are real ({\it e.g.} PBC and ABC), complex functions are
needed for general twist angles. There is a factor of roughly 2.5
in additional CPU time to do the arithmetic to evaluate the
determinant and its derivatives. The actual impact on the total
speed is smaller than this because the calculations of
two-particle quantities such as the potential energy and
correlation factors are still done with real arithmetic; the
actual penalty of working with non-zero twist depends on the
number of particles and the type of the trial wavefunction.
However, as we have discussed earlier, even a factor of 2.5 in
computer time is worthwhile if one is able to approach the
thermodynamic limit quicker, since QMC methods scale as $N^{\nu}$
with $1 \leq \nu \leq 4 $  or, in the case of exact fermion
methods, as $\exp (\gamma N)$. If TABC saves going to larger $N$,
the additional time doing complex arithmetic is well justified.

There are several alternatives for performing the twist averaging:
\begin{enumerate}
\item Evaluate as $\sum_i w_i E_i$ using a grid defined by points
$\bft_i$ with weights $w_i$ (with $\sum_i w_i=1$) in the region
specified by Eq. \ref{unitcell}.
\item Sample the twist during the QMC random walk and take the average.
\item A combination of the two approaches: working on a grid
that is augmented with random displacements.
\end{enumerate}
As we discuss below, all three methods are satisfactory; there is
no fundamental difference in efficiency. The choice of whether to
sample or use a grid is primarily based on convenience and
programming considerations and only secondly on efficiency. We
note that all methods are easy to parallelize.

\vskip .5 cm \noindent{\bf Grid averaging}. First, we must address
the question of which grid and integration rule to use. Since all
properties are periodic with respect to the twist, the grid should
be a Bravais lattice with equally weighted points. One must keep
in mind that the properties, though periodic and continuous, have
discontinuous derivatives at the Bragg planes. Unless grid points
can be located on these planes (which is difficult to achieve in
practice), the integration error will go as
$\epsilon_G\propto\Delta \theta^2\propto N_G^{-2/D}$ where $N_G$
is the number of grid points. Numerically, we find that 16$^3$
grids are needed for an accuracy of $10^{-3}$ (see table I). This
slowly convergent, systematic error is the main drawback of the
grid integration method. For an insulator with a large enough  gap
to excitations, properties would be analytic for all $\bft$ since
the occupation of single particle states will not change as a
function of twist angle and the grid error would converge
exponentially fast.

Once a grid is chosen, one can use symmetry ({\it e.g.} inversion
and rotation through 90$^o$) to reduce the number of grid points
and give them a weight ($w_i$ with $\sum_{i=1}^{N_G} w_i=1$)
proportional to their multiplicity. It is easy to show that the
optimal amount of computer time at each grid point should be
chosen proportional to $w_i/\zeta(\bft_i)^{1/2}$ where the MC
efficiency at $\bft_i$ is defined as $\zeta(\bft_i)=
1/(var(E(\bft_i))cputime)$. We have found on the calculations of
the electron gas described earlier, that this efficiency is
independent of the twist angle except at the special PBC and ABC
points where real functions can be used. Even though we have
symmetry and can integrate over a reduced set of twist values, we
must integrate longer at high multiplicity points since they
contribute more to the average. Hence, the symmetry does not
significantly reduce the needed amount of CPU time.

Since the calculations at different twist angles are uncorrelated,
one can easily show that the efficiency of calculating the twist
averaged energy is given by the relation: \be  \zeta^{-1/2}
=\sum_{i=1}^{N_G} w_i \zeta^{-1/2}(\bft_i).\ee Hence, the overall
efficiency of the TABC energy is an average of the efficiencies of
the individual twist calculations and is higher than that of the
slowest converging twist angle. The additional averaging over
twist angle costs nothing in efficiency.

However, this discussion did not take into consideration start-up
costs at each twist angle, such as the need to reoptimize the
trial wave function at a new twist value, and equilibration costs.
By equilibration, we mean that whenever the twist angle is
changed, enough random walk steps must be taken so that the
configurations are sampled from the new twist value. During this
equilibration, averages cannot be taken. These computational costs
cause a decrease of efficiency by the factor (useful time)/(total
time) and are the main extra computational penalty of the TABC
method within QMC. Since the startup time will scale with the
number of needed grid points, using the above estimate of the
systematic error, $\epsilon_G$, we find that the needed startup
time scales as $\epsilon_G^{-d/2}$ while the time to achieve
equivalent statistical error scales as $\epsilon^{-2}$. Hence, for
very precise calculations $(\epsilon \rightarrow 0)$ and $d<4$,
startup costs can be neglected.

\vskip .5cm \noindent{\bf Twist sampling}. Now consider the second
alternative, where the twist angle is sampled during the random
walk. With this method, we do not have to decide on a grid in
advance and there is no systematic error of a finite grid. Again,
one must equilibrate the configurations after the change of twist
angle and computer time used in that process does not reduce the
variance of the average.

There is an additional increase in variance caused by sampling the
twist angle. One can show that the efficiency decreases by a
factor: $[1+E_{BW}^2/var(E_{\theta})]^{-1}$ where $E_{BW}$ is the
``band-width'' defined in Eq. \ref{bandwidth}. If one spends too
long at a given twist angle, one is not adequately exploring the
twist angle degree of freedom. This gives a definite rule for how
often the twist angle should be updated: the time spent at a given
twist angle should be much longer than the equilibration time but
less than the time needed to get the error in the energy at that
twist value equal to the ``bandwidth'' of the system. If it is not
possible to achieve this relation, the grid scheme should be used.

\vskip .5 cm

With either method, one can achieve more accurate results and less
systematic error by correcting the results using Fermi liquid
theory.  That is, using the twist values and corresponding
energies one can estimate the effective mass and twist averaged
energy using a least squares fit as discussed in Section VI.
However, there could be additional statistical and systematic
error resulting from the fit.

Finally, one can combine the positive features of two methods
using antithetic sampling: use a relatively coarse grid, but then
randomly displace the origin of the grid a number of times during
the run, so as to eliminate the systematic error of the grid and
estimate the true errors. Since the twist angle will change by a
small amount, setup time and equilibration time can be reduced. A
related approach is to sample the twist angle using a quasi-random
number sequence so as to reduce the dispersion of the twist
values.

\end{multicols}
\widetext \vspace*{-5mm}
\begin{multicols}{2}

\end{multicols}

\begin{thebibliography}{10}
\bibitem{multitwist}
Note also that all identical particles must have the same twist.
\end{thebibliography}
\end{document}